\RequirePackage{fix-cm}
\documentclass[twocolumn]{svjour3}          % twocolumn
\smartqed  % flush right qed marks, e.g. at end of proof
\usepackage{graphicx}
\usepackage{adjustbox}

\usepackage{hyperref}
\hypersetup{
    colorlinks=true,
    urlcolor= blue,
    citecolor = black,
    pdftitle={Overleaf Example},
    pdfpagemode=FullScreen,
    }

\usepackage[square,numbers]{natbib}
\usepackage{natbib}
\begin{document}

\title{Open and free EEG datasets for epilepsy diagnosis %\thanks{Grants or other notes
}
\subtitle{}


\author{Palak Handa  \and  Monika Mathur     \and
        Nidhi Goel   
}

\institute{P. Handa \at
              Dept. of ECE, DTU, Delhi \\
              \email{palakhanda97@gmail.com}           %  \\
           \and
           M. Mathur \at
              Dept. of ECE, IGDTUW, Delhi  \\   
              \email{mathur.monika2007@gmail.com}
              \and
          N. Goel \at
              Dept. of ECE, IGDTUW, Delhi  \\   
              \email{nidhi.iitr1@gmail.com}}

\date{Received: date / Accepted: date}

\maketitle

\begin{abstract}
The Epilepsies are a common, chronic neurological disorder affecting more than 50 million individuals across the globe. 
It is characterized by unprovoked, recurring (similar or different type) seizures which are commonly diagnosed through clinical EEGs. Good-quality, open-access and free EEG data can act as a catalyst for on-going state-of-the-art (SOTA) research works for detection, prediction and management of epilepsy and seizures. They can also aid in improving the quality of life (QOL) of these diseased individuals and contribute research in healthcare multimedia, data analytics and Artificial Intelligence (AI) in personalized medicine. This paper presents widely used, available, open and free EEG datasets available for epilepsy and seizure diagnosis. A brief comparison and discussion of open and private datasets has also been done. Such datasets will help in development and evaluation of automatic computer-aided system in healthcare. 

\end{abstract}

\section{Introduction}
\label{intro}

The Epilepsies are a chronic neurological disorder characterized by unprovoked, recurring (similar or different type) seizures. A clinical EEG setting is used by doctors to observe different types of epileptic activity as it leaves distinct impressions in the form of interictal epileptiform discharges, peri-ictal activities and high frequency oscillations \emph{etc} \cite{thomas2001economic}.

The annual economic burden of epilepsy is enormous in developing countries like India where it is estimated to be 88.2\% of gross national product (GNP) per capita and 0.5\% of the overall GNP \cite{thomas2001economic}. Hence, early diagnosis through recent technologies like Artificial Intelligence (AI), feature engineering, data analytics and multimedia is vital and can aid in Quality Of Life (QOF) of patients and their associated caretakers.  

Secured, reproducible AI algorithm, good quality data and efficient computing horse power are the major elements for development of early detection and prediction of epileptic wave forms through EEG signals. There are several types of EEGs such as intracranial, scalp, ambulatory, \emph{etc.} They are recorded in video, image and signal format depending on their use and application in hospitals. 

There is a huge demand of real-time biomedical multimedia tools for data analysis, and pattern recognition of such formats. Bonn EEG time series database \cite{andrzejak2001indications} was the first EEG dataset to be publically available for research applications in this field. It remains as the benchmark dataset for most research works due to its availability. Several datasets discussed in this paper encourage scientific advances in this field. 

The data quality of biomedical datasets is measured through various factors such as presence of artefacts and noise, missing values, descriptive information, annotations by health experts, pre-fined data structure, processing and robustness to outliers \emph{etc.}

The datasets mentioned in the section 2 are freely available (except temple university which requires a login), re-distributable for research purposes, previously freely available and or became private or removed from the portal based on adult and paediatric population. Table 1 shows a comparison of publically available, open-access (except temple university which requires a login) human EEG datasets for epilepsy diagnosis. 

The source and availability of these were verified on 26-07-2021, which may change in the future. They were found using different keywords like `EEG datasets for epilepsy', `datasets for epilepsy detection', `EEG based epilepsy diagnosis', and `open EEG datasets' on Pub med and google scholar search engine.

\section{EEG datasets for epilepsy diagnosis}
\label{sec:1}

There are several EEG datasets for epilepsy diagnosis which are freely available and private due to various reasons such as lack of ethical clearance. This section lists all the existing EEG datasets with their URLs for adult and paediatric population.

\subsection{Adult datasets}

The adult population consists of affected individuals above 20 years of age. There are several databases like American Epilepsy Society Seizure Prediction Challenge database \cite{howbert2014forecasting}, data\-set of EEG recordings of pediatric patients with epilepsy based on the 10-20 system \cite{ds003555:1.0.1} and Karunya University \cite{selvaraj2014eeg}
which contain both adult and paediatric EEG data. They are mentioned in section 2.2.

\subsubsection{Bonn EEG time series database\cite{andrzejak2001indications}}

This database comprises of 100 single channels EEG 
of 23.6 seconds with sampling rate of 173.61 Hz. 
Its spectral bandwidth range is between 0.5 Hz and 85 Hz. 
It was taken from a 128 channel acquisition system. Five patients EEG sets were cut out from a multi-channel EEG recording and named A, B, C, D and E. Set A and B are the surface EEG recorded during eyes closed and open situation of healthy patients respectively. Set C and D are the intracranial EEG recorded during a seizure free from within seizure generating area and from outside seizure generating area of epileptic patients respectively. 
Set E is the intracranial EEG of an epileptic patient during epileptic seizures. Each set contains 100 text files wherein each text file has 4097 samples of 1 EEG time series in ASCII code. A band pass filter with cut off frequency as 0.53 Hz and 40 Hz has been applied on the data.  
It is an artifact free data and hence no prior pre-processing is required for the classification of healthy (non-epileptic) and un-healthy (epileptic) signals. The strong eye movement’s artefacts were omitted. It was made available in 2001. 
The extended version of this data is now a part of EPILEPSIA project.
Available link: \href{https://repositori.upf.edu/handle/10230/42894}{a} and
\href{https://www.ukbonn.org/epileptologie/ag-lehnertz-downloads/}{b} 

\subsubsection{Bern-Barcelona EEG database \cite{andrzejak2012nonrandomness}}

This multi channel EEG database was recorded using specialized electrodes and consists of five patients with longstanding pharmacoresistant temporal lobe epilepsy. The patients underwent epilepsy surgery. The sampling rate was either 512 or 1024 Hz based upon whether they were recorded with less or more than 64 channels of EEG system. Three out of five attained complete seizure freedom. Two types of EEG are present in the data \emph{i.e.,} focal and non-focal. Each file has about 10240 samples for a time duration of 20 seconds. 
Available link: \href{https://repositori.upf.edu/handle/10230/42829}{c} and
\href{https://www.upf.edu/web/ntsa/downloads/-/asset\_publisher/xvT6E4pczrBw/content/2012-nonrandomness-nonlinear-dependence-and-nonstationarity-of-electroencephalographic-recordings-from-epilepsy-patients#.YQJnBI4zbIU}{d}

\subsubsection{Temple University EEG corpus \cite{obeid2016temple}}

Temple University EEG corpus is the largest free EEG data available for epilepsy and seizure types diagnosis till date. It consists of data acquired from 2000 to 2013 using different EEG clinical settings for about 10,874 patients. This community has developed various software products such as annotation tools, toolboxes for seizure detection, and EDF browser for data analysis of EEG, EMG, and ECG \emph{etc} signals. EDF browser helps to view EEG recording in a video form. There are various datasets available such as IBM Features For Seizure Detection (IBMFT), the TUH EEG epilepsy corpus, seizure corpus, slowing corpus, and events corpus \emph{etc}. A user ID and password is required to get access to these datasets. \href{https://www.isip.piconepress.com/projects/tuh\_eeg/html/downloads.shtml}{Available link.}

\subsubsection{Neurology and sleep center, New Delhi EEG dataset \cite{swami2016eeg}}
This database comprises of 5.12 seconds EEG data. It was recorded using 57 EEG channel Grass Tele-factor Comet AS40 Amplification System; sampled at 200 Hz. It’s spectral bandwidth range is between 0.5 Hz and 70 Hz. Time series EEG datasets are categorized into three major MATLAB file folder namely ictal, pre-ictal and inter-ictal stages. Each MAT file has 1024 samples. A subset of this database is publically available. \href{https://www.researchgate.net/publication/308719109_EEG_Epilepsy\_Datasets}{Available link.}

\subsubsection{Epileptic Seizure Recognition Data Set \cite{andrzejak2001indications}}
The time series EEG dataset consists of 11500 instances of EEGs
of 4 subjects suffering from epilepsy. 
This data has been removed from the UCI machine learning repository recently and was released in 2017.
It is a simplified version of the original data released by \cite{andrzejak2001indications}. It consists of 5 subjects (4 unhealthy and 1 healthy) 
performing different activities and experience epileptic seizures except subject 1. The time duration for each EEG was 23.5 seconds. 

\subsubsection{Siena Scalp EEG Database \cite{detti2020eeg, ss1}}

This multi-channel EEG database of 14 epileptic patients (9 males and 5 females) was recorded using specialized amplifiers, and reusable electrodes. The signals were recorded with a sampling rate of 512 Hz and stored in EDF files. The data has been acquired from Unit of Neurology and Neurophysiology of the University of Siena, Italy and focuses on seizure prediction. It is an integral part of national interdisciplinary research project PANACEE. This data contains 47 seizures from 128 hours of video EEG recording. The start and end time of a seizure was also recorded and contains the list of electrodes present on the scalp of a patient during event recording. Three types of seizures namely focal onset with and without impaired awareness, and focal to bilateral
tonic–clonic (FBTC) were found and recorded in the diseased patients.  \href{https://physionet.org/content/siena-scalp-eeg/1.0.0/
}{Available link.}

\subsubsection{Single electrode EEG data of healthy and epileptic patients \cite{panwar2019automated, panwar_siddharth_2020_3684992}}
This dataset was generated with a motive to build predictive epilepsy diagnosis model and publically available since 2020. It was generated on a similar acquisition and settings \emph{i.e.,} sampling frequency, bandpass filtering and number of signals and time duration as of University of Bonn. It has overcome the limitations faced by University of Bonn dataset such as different EEG recording (inter-cranial and scalp) for healthy and epileptic patients \cite{panwar2019automated}. All the data were taken exclusively using surface EEG electrodes for 15 healthy and epileptic patients.  \href{https://zenodo.org/record/3684992#.YQJod44zbIU}{Available link.}

\subsubsection{Epileptic EEG Dataset \cite{mendeleydata}}
This multi-channel, long term EEG database was 
record\-ed for 6 patients suffering from focal epilepsy. They were undergoing pre-surgical evaluation for possible epilepsy surgery. 
Different EEG segments of a seizure like ictal, pre-ictal, inter-ictal and its onsets have been included in the data. The signals were recorded with a sampling rate of 500 Hz and stored in EDF files. 
Labelled and classified data points (train and test set) have been mentioned for complex partial electrographic, and video-detected seizures. All the EEG signals underwent band pass filtering of range 1-70 Hz where 50 Hz (utility frequency) was also removed. \href{https://data.mendeley.com/datasets/5pc2j46cbc/1}{Available link.}

\subsection{Paediatric datasets}

The paediatric EEG database consists of affected individuals from age 1 month - 20 years. 

\subsubsection{Children’s hospital Boston–MIT database \cite{goldberger2000physiobank}}

This database comprises of 844 hour continuous EEG. 23 pediatric patients from age 1.5-19 who underwent scalp multi-channel EEG recording. It is the first paediatric EEG database available for epilepsy and seizure diagnosis. 
The patients were given anti-seizure medications. About 200 seizures were recorded in a universal bio-polar montage with about 24-27 EEG channels. 
Sampling frequency was kept to be 256 Hz.
Each EEG segment is called as a record which usually is for duration of one hour. There are 9-42 edf files from a single subject. 
Additional vagal nerve stimulus signals are also present. Separate file name and montages have been mentioned for seizure v/s non seizure episode in EEG segments. \href{https://physionet.org/content/chbmit/1.0.0/}{Available link.}

\subsubsection{Karunya University \cite{selvaraj2014eeg}}

This database comprises of 18 channel EEG data with 
segments of normal, focal and generalized epileptic se\-izure 
activities from 1–107 years of patients. It was released in 2014 but the website is not available for research use now. 
Each segment has 2056 sample points. Sampling frequency was kept to be 256 Hz. The EEG recordings vary from 40 minutes to one hour.
It was collected from a diagnostic center based in Coimbatore, India. \href{http://eegdb.byethost7.com/public/}{Available link.} 

\subsubsection{A dataset of neonatal EEG recordings with seizures annotations \cite{nathan_stevenson_2018_2547147}}

This database consists of multi-channel, good quality EEG recordings of 79 term neonates where 39 of them suffered from neonatal seizures in the NICU of Helsinki University Hospital, Finland. The recordings were captured with NicOne EEG amplifier, and 19 EEG channel cap. 
The signals were recorded with a sampling rate of 256 Hz and stored in EDF files. It consists of seizure annotations by healthcare experts for seizure detection purpose. The data was pre-processed using butterworth high-pass filtering. The data also contains natural artefacts. \href{https://zenodo.org/record/1280684#.YQJpgI4zbIU}{Available link.}

\begin{table*}
\centering
\caption{Comparison of existing EEG datasets for epilepsy diagnosis}
\label{tab:1}       % Give a unique label
\begin{adjustbox}{angle=90}
\centering
\begin{tabular}{llllllllll}
\hline
Ref. & Availability & Type & Source & Year & Size & No. of & No. of & Sampling  & EEG segments  \\ 
&&&& &&channels&patients & frequency & \\ \hline 
\cite{andrzejak2001indications} & Freely available & Adult & e-repositori upf. & 2001 & 3.05 MB & 100 single & 5 & 173.61 Hz & seizure states, healthy \\
\cite{goldberger2000physiobank} & Freely available & Paediatric & PhysioNet repository  & 2010 &  40 GB & 23-26 & 22 & 256 Hz  & Intractable seizures \\

\cite{andrzejak2012nonrandomness} & Freely available & Adult & e-repositori upf.  & 2012 & 814 MB & 64  & 5 & 512 Hz & Focal, Non-focal \\

\cite{howbert2014forecasting} & Freely available & Dog and human & Kaggle & 2014 & 105 GB & - & - & - & different types \\

\cite{selvaraj2014eeg} & Not available & Adult and Paediatric & Website & 2014 & - & - & - & 256 Hz & normal, focal   \\
&& & & & & && & and generalized  \\
&& & & & & && &epileptic seizures \\
\cite{obeid2016temple}& Free but   & Adult & Website  & 2015 & 572 GB  & 20-31 & 10,874 & 250,  & different types \\
& requires login && & &&& & 256, 512 Hz & \\ 

\cite{swami2016eeg} & Freely available & Adult & Researchgate &  2016 & 604 KB & 57 & 10  & 200 Hz  & Ictal, inter-ictal, \\
&& & & & & && & pre-ictal EEGs \\

\cite{andrzejak2001indications} & Removed & Adult & UCI repository & 2017 & 3 MB & 100 single & 5 & 173.61 Hz & seizure related, healthy \\

\cite{nathan_stevenson_2018_2547147}& Freely available & Paediatric (neonates)  & Zenedo  & 2018 &  4.3 GB & 19 & 79 & 256 Hz & seizure onset \\

\cite{raghu2019performance} & Private & Adult & - & 2019 & - & 19 & 115 & 128 Hz & epileptic and healthy \\
\cite{panwar2019automated} & Private & Adult & - & 2019 & - & - & 50 & 250, 256 Hz & generalized and  \\
&& & & & & && &focal epilepsies \\

\cite{wu2019automatic} & Private & Adult & - &2019  & - & 21  & 5 & 500 Hz  & focal and \\
&& & & & & && &tonic-clonic \\

\cite{avcu2019seizure}  & Private & Pediatric & - & 2019 & - & - &29& 200, 500 Hz & typical absence seizures\\
\cite{yan2019automated}  & Private & Adult & - & 2019 & - & - & 12 & 256 Hz & seizure events \\
\cite{choi2019novel} & Private & - & - & 2019 & - & 21 & 25 & 200 Hz & seizure events \\
\cite{cao2019epileptic} & Private & - & - & 2019 & - & 18 & 10 & 256 Hz & seizure states \\
\cite{bilal2019automatic} & Private & - & - & 2019 & - & 22 & 22 & 250 Hz & ictal, non-ictal \\
\cite{panwar_siddharth_2020_3684992}& Freely available & Adult & Zenedo & 2020 &  20 MB & - & 15 & 173.61 Hz & Inter-ictal \\

\cite{yedurkar2020multiresolution} & Private & - & - & 2020 & - & 21 & - & 250 Hz & seizure onsets \\
\cite{das2020epileptic} & Private & Adult & - & 2020 & - & 21 & 150 & 256 Hz & seizure and normal \\ 
\cite{ss1}& Freely available & Adult & PhysioNet repository & 2020 & 20 GB & 29 & 14 & 512 Hz  & Epileptic seizures    \\
&&  & && && & & (focal onset, tonic-clonic)\\

\cite{mendeleydata}& Freely available & Adult & Mendeley repository  & 2021 &  3133 MB & 21 & 6 & 500 Hz & Complex partial, \\
&& && & & && & electrographic and \\
&& && & && & &video-detected seizures \\

\cite{ds003555:1.0.1}& Freely available & Paediatric and Adult & Open neuro repository & 2021 &  15 GB  & 52 & 30 & 2000 Hz & HFO markings \\ \hline

\end{tabular}
\end{adjustbox}
\end{table*}

\subsubsection{Dataset of EEG recordings of pediatric patients with epilepsy based on the 10-20 system \cite{ds003555:1.0.1}}

This dataset consists of scalp EEG recordings to study the impact age on observed High Frequency Oscillations (HFO) in pediatric epileptic patients. Three hours of pediatric and adult EEG sleep data was recorded for 30 focal or generalised epileptic patients. The signals were recorded with a sampling rate of 2000 Hz and stored in EDF files. Different sleep stage annotations are available in this database. \href{https://openneuro.org/datasets/ds003555/versions/1.0.1U}{Available link.}

\subsection {Others}
\subsubsection{American Epilepsy Society Seizure Prediction Challenge database \cite{howbert2014forecasting}}

This database consists of intra-cranial EEG segments from dogs and humans with different acquisitions of sampling rate, duration of EEGs, and no. of electrodes \emph{etc.} It was released as a part of the kaggle challenge hosted by the American Epilepsy Society in 2014 for development of seizure forecasting systems and witnessed about 504 teams. Different seizure segments of ictal, pre-ictal, post-ictal, inter-ictal were provided in MATLAB files. The data storage was about 105 GB. Additional annotated EEG data was also provided by the University of Pennsylvania and the Mayo Clinic. \href{https://www.kaggle.com/c/seizure-prediction/data}{Available link.} 

\subsubsection{Private databases}
Several private databases have also been recorded for epilepsy diagnosis using EEG signals \cite{panwar2019automated, das2020epileptic, raghu2019performance, wu2019automatic, avcu2019seizure, bilal2019automatic, yedurkar2020multiresolution, yan2019automated, choi2019novel, cao2019epileptic}. The European Epilepsy database \cite{ihle2012epilepsiae} is a private database which consists of high quality, annotated EEG signals from University of Bonn \cite{andrzejak2001indications}, Freiburg \cite{schelter2006false}, Flint hills, and many multi-modal like MRI imaging data. The website of \cite{selvaraj2014eeg} is not available. EEG data in \cite{zwolinski2010open} was freely available till 2015 when its portal crashed.

\section{Conclusion}
Diagnosis, treatment and management of epilepsy is still a challenging task for the scientific and healthcare community. It’s detection by visual introspection of long hour EEG is not only time taking but a very tedious and subjective task. Artificial intelligence can help in escalating this process and lead to successful detection of different types of epilepsies through efficient, high quality and annotated EEG data. This paper has presented all the existing EEG datasets for epilepsy diagnosis with its availability and brief comparisons. Such datasets motivate scientific research in early diagnosis of epilepsy through robust techniques.

%\begin{acknowledgements}
%\end{acknowledgements}

\section*{Conflict of interest}
The authors declare that they have no conflict of interest.


\bibliography{sample}
\end{document}